\documentclass[onecolumn,showpacs,
superscriptaddress,10pt,preprint,aps,prb]{revtex4-1}

\usepackage{graphicx}% Include figure files 
\usepackage{dcolumn}% Align table columns on decimal point 
\usepackage{bm}% bold math 
\usepackage{amsmath} 
\usepackage{epstopdf}
\AppendGraphicsExtensions{.tif} 
\usepackage{epsfig} 
\usepackage{subfigure} 
\usepackage{fancyvrb} 
\usepackage[usenames,dvipsnames]{color} % may cancel this after it is all done 
\usepackage[normalem]{ulem} % for special text properties. may cancel. 
\usepackage{tikz} % for drawing graphics 
\usepackage[english]{babel} 
 
\newcount\colveccount
\newcommand*\colvec[1]{
        \global\colveccount#1
        \begin{pmatrix}
        \colvecnext
}
\def\colvecnext#1{
        #1
        \global\advance\colveccount-1
        \ifnum\colveccount>0
                \\
                \expandafter\colvecnext
        \else
                \end{pmatrix}
        \fi
}
\begin{document}

\title[Selective Radiative Heating of Nanostructures Using Hyperbolic Metamaterials]{Selective Radiative Heating of Nanostructures Using Hyperbolic Metamaterials}% Force line breaks with \\
%\thanks{Footnote to title of article.}

\author{D. Ding}
\author{A. J. Minnich}%
 \email{aminnich@caltech.edu}
 \noaffiliation
\affiliation{ 
Division of Engineering and Applied Science, California Institute of Technology, Pasadena, California 91125, USA}%

\date{\today}% It is always \today, today,

\begin{abstract}
Hyperbolic metamaterials (HMM) are of great interest due to their ability to break the diffraction limit for imaging and enhance near-field radiative heat transfer. Here we demonstrate that an annular, transparent HMM enables selective heating of a sub-wavelength plasmonic nanowire by controlling the angular mode number of a plasmonic resonance. A nanowire emitter, surrounded by an HMM, appears dark to incoming radiation from an adjacent nanowire emitter unless the second emitter is surrounded by an identical lens such that the wavelength and angular mode of the plasmonic resonance match. Our result can find applications in radiative thermal management.  
\end{abstract}

\maketitle

Engineering thermal radiation is of importance for a number of technologies, including infrared imaging, energy conversion, thermal
insulation, thermal signature control, and thermal management \cite{Liu2014}. Recent works have demonstrated that far-field spectral and angular characteristics of thermal radiation can be controlled using photonic crystals \cite{Lee2007,Han2010,Yeng2012} and metamaterials \cite{Liu2011,Bermel2011,Mattiucci2012,Wang2013}. These structures can also enable near-field resonant surface modes to propagate into the far-field using gratings \cite{Greffet2002} and antennas \cite{Schuller2009} to out-couple surface modes. In the near field, radiative heat transfer can be greatly enhanced due to the presence of evanescent waves \cite{Cravalho1967,Polder1971}. These enhancements have recently been demonstrated experimentally \cite{Kittel2005,Shen2009,Rousseau2009,Ottens2011}. Thermal radiation into the far-field can also be enhanced in a thermal extraction scheme in which an impedance-matched extraction device allows the propagation of internally reflected modes \cite{Yu2013}. 

Recently, hyperbolic metamaterials (HMMs) have been under intense investigation for their potential to control thermal radiation. HMMs possess dielectric constants of opposite sign along different axes and hence allow the propagation of high momentum modes within the HMM due to the hyperbolic dispersion \cite{Smith2003}. HMMs can be fabricated in practice is as a multilayer stack with alternating materials of opposite sign of dielectric constant. HMMs were originally of interest for their potential to project images with resolution below the diffraction limit into the far-field, as proposed theoretically \cite{Jacob2006,Salandrino2006} and later demonstrated experimentally \cite{Liu2007}. For thermal radiation, HMMs have been studied for their potential to enhance near-field heat transfer \cite{Biehs2012,Guo2013,Liu2013,Guo2014} as well as control the spectral and angular distribution of far-field radiation \cite{Narimanov2011,Molesky2013,Ji2014,Nefedov2014}.

Here, we examine how HMMs modify the far-field thermal emission spectrum of nanostructures. We find that a lossy plasmonic nanowire surrounded by a transparent, annular HMM lens yields thermal emission that primarily occurs only at a specific wavelength and angular mode number and greatly exceeds that of the nanowire alone. This angular mode resonance enables highly selective radiative heating because only nanowires that are surrounded by identical HMM lenses can exchange radiation.

\begin{figure}[h!]
\centering
    \includegraphics[width=1\textwidth]{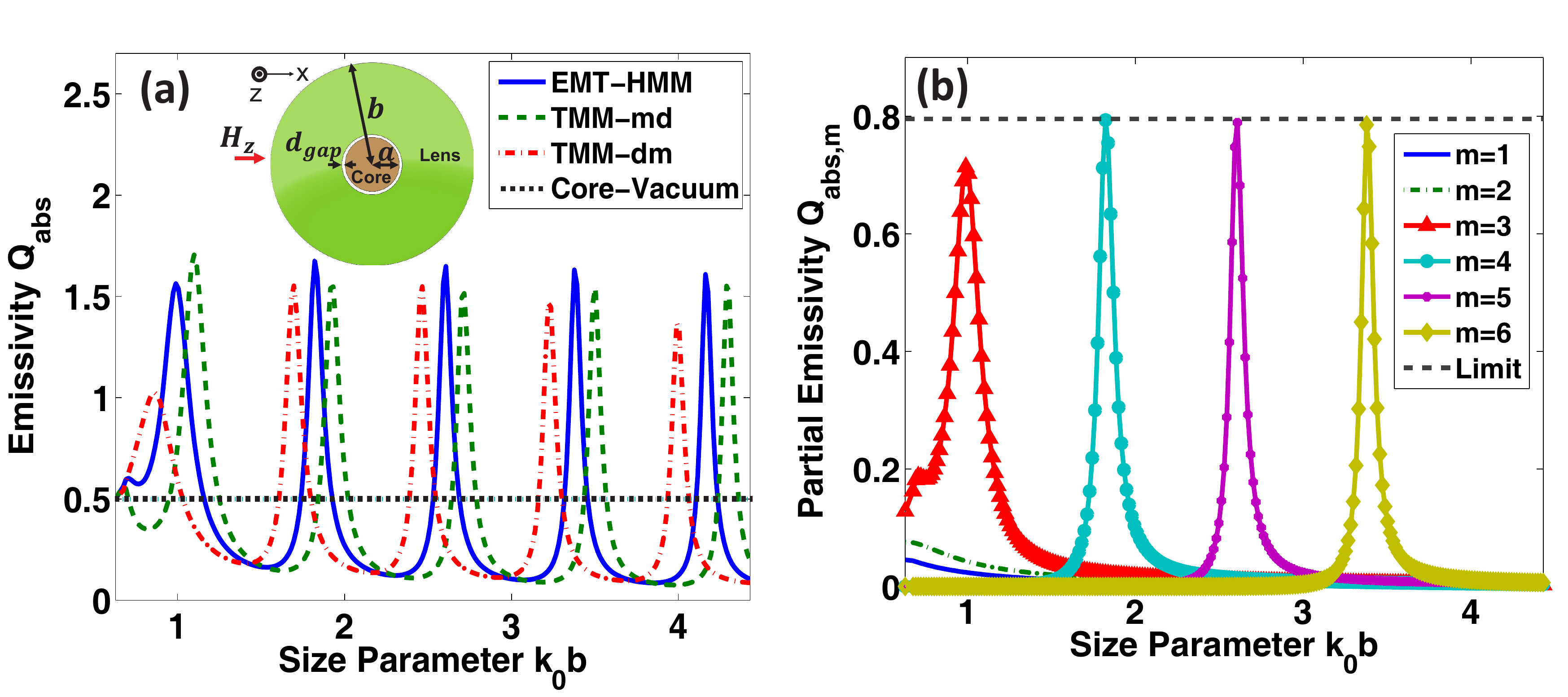}
\caption{\label{fig:mode_analyze} (a) Absorption efficiency, or equivalently emissivity, versus size of lens $b$ for core size of $a=0.1\lambda$ with different lenses surrounding a plasmonic core. Core-Vacuum (black dotted line) indicates $Q_{abs}$ of only a core of size $a$. $Q_{abs}$ for the core-HMM lens calculated using EMT, TMM-md and TMM-dm are shown as the dark blue solid, green dashed and the red dotted-dashed line, respectively.  There are many resonant peaks that enhance the emissivity over that of the bare core when the HMM lens is present. Inset: Schematic of the geometry. The core and the lens have radius $a$ and $b$, respectively. (b) Partial contribution to total absorption efficiency for each angular mode $m$ in (a). The dashed black line is the single-channel limit defined in the text. The mode $m=4$ achieves the single channel limit, unlike $m=3$.}
\end{figure}

% Describe TMM and system setup
We begin by considering a lossy nanowire core of radius $a$ that is in optical contact with a lens medium in vacuum, as shown in 
the inset of Fig.~\ref{fig:mode_analyze}(a). The system is assumed to be infinite in the z direction with the polarization such that $\mathbf{E}\perp z$. The magnetic field $H_z(\vec{r})$ from an incident plane wave can be expressed as
\begin{widetext}
 \begin{align}
     H_z(\vec{r})&= \left\{
     \begin{array}{lr}
       \sum^\infty_{m=-\infty} (i)^m \left(J_m(k_0 \rho)-a_m H^{(1)}_m(k_0 \rho)\right)                     \exp(i m \phi):& \rho>b\\
       \sum^\infty_{m=-\infty} (i)^m \left(c^{(j)}_m J_m(k_j \rho)+d^{(j)}_m H^{(1)}_m(k_j \rho)\right)                     \exp(i m \phi):& a<\rho<b\\
       \sum^\infty_{m=-\infty} (i)^m b_m J_m(k_j \rho)                     \exp(i m \phi):& \rho<a
     \end{array}
   \right. \label{Eq:field}
\end{align}
\end{widetext}
where $J_m$ and $H^{(1)}_m$ are Bessel and Hankel functions of the first kind of angular modal number $m$ in cylindrical coordinates \cite{Scheuer2005} and $a_m$ and $b_m$ are the coefficients of the Hankel function of the scattered field in outermost vacuum ($\rho>b$) and the Bessel function of the transmitted field of core ($\rho<a$), respectively. $c^{(j)}_m$ and $d^{(j)}_m$ are the coefficients for the Bessel and Hankel function of the field in the $j$th layer, and $k_j$ denotes the wave vector of each $j$th layer up to the core. $k_0$ denotes the wave vector in vacuum. The coefficients $a_m$, $b_m$, $c^{(j)}_m$ and $d^{(j)}_m$ can be solved by matching boundary conditions of tangential fields at the boundary of each layer using the Transfer Matrix Method (TMM) in cylindrical coordinates \cite{Chew1999,Nikolaev1999,Scheuer2005}. In this method, the continuity of the tangential components of the $\mathbf{E}$ and $\mathbf{H}$ fields at the boundary of each layer derived from Equation \ref{Eq:field} can be written in matrix form and the transfer matrix of the whole system can be calculated by multiplying the transfer matrices of individual interfaces. 

The absorption efficiency $Q_{abs}$ can then be expressed as
\begin{equation} 
Q_{abs}=\sum^\infty_{m=-\infty}Q_{abs,m}=\frac{2}{k_0 a}\sum^\infty_{m=-\infty} \operatorname{Re}(a_m)-\lvert a_m \rvert ^2 
\label{Eq:Qabs}
\end{equation}
where $Q_{abs,m}$ is the partial absorption efficiency or absorption efficiency per mode and $\operatorname{Re}(a_m)$ is the real part of the coefficient for mode $m$ defined in Equation \ref{Eq:field} according to Mie theory \cite{Hulst1981,Bohren1998}. By Kirchoff's law, the absorptivity equals the emissivity for each direction and wavelength \cite{Modest2003}, and hence $Q_{abs}$ can be interpreted as the emissivity. Note that the emissivity can exceed unity for subwavelength objects because the absorption cross-section can be larger than the geometric cross-section \cite{Bohren1998}.

The HMM lens consists of an alternating layered structure of dielectric and metal leading to anisotropic permittivity along the radial and tangential direction. These anisotropic dielectric constants can be expressed using effective medium theory (EMT) according to $(\epsilon_{\rho},\epsilon_{\theta})=\left(\epsilon_m\epsilon_d /\left((1-f)\epsilon_m+f\epsilon_d\right),f\epsilon_m+(1-f)\epsilon_d \right)$ where $f$ is the volume fraction occupied by the metal and $\epsilon_m$, $\epsilon_d$ are the respective metal and dielectric permittivities \cite{Jacob2006}. We assume the lens to be lossless and define $Q_{abs}$ with respect to the core radius $a$. Neglecting loss in the lens means that the lens cannot exchange thermal radiation with the core, and thus its contribution to heat transfer can be neglected.  

% Core-Vacuum Calculation
First, we consider the emissive properties of only the nanowire core of radius $a$ in vacuum without any lens. Figure \ref{fig:mode_analyze}(a) shows the computed emissivity $Q_{abs}$ for the nanowire core with a permittivity of $-1.05+0.01i$ and  $a=0.1\lambda$ where $\lambda$ is the wavelength of the incident field. We choose the core permittivity close to the ideal plasmonic resonance condition to demonstrate our result but other negative real permittivity values can be chosen with similar results. We assume a typical wavelength $\lambda=10\ \mu$m, corresponding to the maximum of the blackbody spectrum around 290 K, giving $a=1\ \mu$m and yielding an emissivity of 0.5. The maximum emissivities for the nanowire core decrease with increasing size parameter as the absorption efficiency scales as $1/a$. Note that plasmonic resonances do occur at specific sizes for a given permittivity for a nanowire core \cite{Lukyanchuk2006}, but tuning the angular mode number of the resonance requires changing the permittivity of the nanowire. 

% Result of metal or dielectric lens
Now, consider the nanowire surrounded by a transparent material called ``lens" as shown in inset of Fig.~\ref{fig:mode_analyze}(a). The transparent lens is assumed to be lossless such that it cannot exchange radiation with the core. The total thickness of the core and lens $b$ considered in Fig.~\ref{fig:mode_analyze}(a) ranges from $1-7\ \mu$m with corresponding size parameters $k_0b$ shown. We assume a vacuum gap of width $\lambda/200$ (50 nm for $\lambda=10\ \mu$m) exists to prevent heat conduction, although this assumption does not affect our conclusions. The addition of this lens with a lossless metal of permittivity $\epsilon=-1.05$ or a dielectric of permittivity $\epsilon=10$ results in a lower emissivity $Q_{abs}$ than the bare core (Core-Vacuum) case. This reduction in emissivity can be attributed to the impedance mismatch between the lens and vacuum that reflects some modes before they reach the absorptive core.

% method to simulate HMM lens
Next, consider the nanowire surrounded by a transparent HMM lens. We compute $a_m$ in this case using either EMT or considering each individual layer of the HMM with a transfer matrix. For the EMT-HMM case, we scale $m$ in Equation \ref{Eq:field} of the HMM layer \cite{Ni2010} to $m'=m\sqrt{\epsilon_{\theta}/\epsilon_{\rho}}$. For the layer by layer case, the thickness of each metal-dielectric bi-layer is chosen to be $\lambda/400$ (25 nm for $\lambda=10\ \mu$m). We examine both the metal-dielectric (TMM-md) and dielectric-metal (TMM-dm) structures such that the first layer adjacent to the core is a metal or a dielectric, respectively. For the EMT-HMM case, we take the optical constants to be $(\epsilon_{\rho},\epsilon_{\theta})=(10,-0.025)$ according to EMT. For the TMM calculation, we take $(\epsilon_m,\epsilon_d)=(-5.1,3.4)$ with $f=0.4$, giving the same values of $(\epsilon_m,\epsilon_d)=(10,-0.025)$ as EMT.

This calculation is plotted in Fig.~\ref{fig:mode_analyze}(a). In contrast to decrease of emissivity $Q_{abs}$ with the metal and dielectric lens, the emissivity $Q_{abs}$ with the HMM lens exhibits strong peaks as the size parameter increases for both the EMT and TMM calculations. The emissivity $Q_{abs}$ peaks in the TMM-md and TMM-dm cases are in close proximity to the right and left of the EMT-HMM peaks, respectively, and converge to the EMT result as the layer thickness decreases. Thus, by placing a HMM lens of the right size at one of these peaks around the core, the emissivity can be increased by about three times compared to the same bare nanowire core. Larger enhancements of 4-5 times relative to the bare core can be achieved at larger core sizes for the same loss of the core. Enhancements greater than 50 times that of a larger bare core can be achieved if the loss of the core is optimized but the required small loss is not realistic for any available plasmonic materials and thus is not considered further. 

To understand the origin of these peaks, we examine the decomposition of absorptivity from the EMT-HMM case in Fig.~\ref{fig:mode_analyze}(a) into partial absorptivity for modes $m=1$ to $m=6$ as shown in Fig.~\ref{fig:mode_analyze}(b). The $m=1$ and $m=2$ cases do not have resonant peaks for the given size range but modes $m=3$ to $m=6$ each have a specific resonance at different size parameters $k_0b$.  These resonant size parameters correspond to the same peak positions in Fig.~\ref{fig:mode_analyze}(a) and achieve emissivity close to the well-known single channel limit \cite{Schuller2009}. At a given size parameter, most of the total absorption cross-section is due to a single resonant angular mode.

\begin{figure}[h!]
\centering
%\begin{subfigure}
% \centering
    \includegraphics[width=1\textwidth]{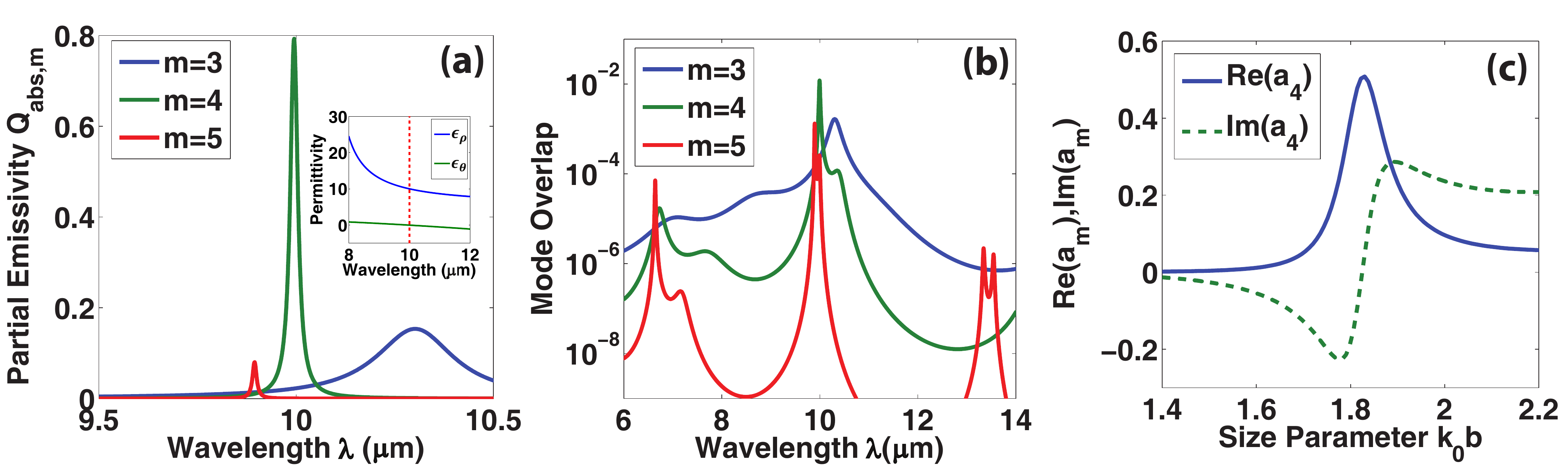}
%\end{subfigure}
\caption{\label{fig:wavelength_dept} (a) Partial emissivity versus wavelength assuming that all optical properties follow a Drude model. Only a few angular modes contribute to radiative transfer at specific wavelengths. Inset: relative permittivities $(\epsilon_{\rho},\epsilon_{\theta})$ of the HMM lens for the range of wavelengths considered. The red dashed line at $10\mu$m indicates the permittivities used in Fig.~\ref{fig:mode_analyze}. (b) Product of partial emissivity $Q_{abs,m}$, as in (a), versus wavelength for two size parameters $k_0 b=2.6$ and $k_0 b=1.8$ for the modes $m=3,4,5$. There is very little overlap of all modes as two systems do not share an angular mode resonance. (c) Real and imaginary part of $a_m$ defined in Equation \ref{Eq:field} for $m=4$ in Fig.~\ref{fig:mode_analyze}. This angular mode satisfies the condition for single-channel limit at the chosen size parameter of $k_0b=1.8$.}
\end{figure}

Further, assuming a Drude model for optical properties, this resonance yields by far the largest emissivity over a considerable range of wavelength. We examine the wavelength dependence of the enhancement in thermal emission that can be achieved using the HMM lens using a Drude model given by $\epsilon_{m}=1-\omega_p^2/(\omega^2 + i \gamma\omega)$, where $\omega$ is the frequency and $\omega_p$ is the plasmon frequency. For the core, $\gamma=0.0035\omega_p$ and $\lambda_p=2\pi c/\omega_p=7\ \mu$m. The metal in the HMM is assumed to have a Drude dispersion that is lossless ($\gamma=0$) and $\lambda_p=4.05\ \mu$m. These parameters yield the same permittivities as used in Fig.~\ref{fig:mode_analyze} at a wavelength $\lambda=10\ \mu$m as shown in the inset in Fig.~\ref{fig:wavelength_dept}(a). The partial emissivity $Q_{abs,m}$ versus wavelength for size parameter $k_0b=1.8$, at the resonance for $m=4$, is plotted in Fig.~\ref{fig:wavelength_dept}(a). At a particular wavelength, the emissivity is nearly entirely due to a single angular mode; for example, the resonant peak at $10\ \mu$m is nearly completely due to $m=4$ mode, with a small additional contribution from $m=3$ but not from $m=5$. 

We now compare the overlap of these resonances for identical nanowires surrounded by HMM lenses of different size parameters by multiplying the partial emissivity $Q_{abs,m}$ from Fig.~\ref{fig:wavelength_dept}(a) for two different size parameters, $k_0 b=2.6$ and $k_0 b=1.8$, for modes $m=3,4,5$. As shown in Fig.~\ref{fig:wavelength_dept}(b), there is negligible overlap between the partial emissivity of the two cases over the full range of the blackbody spectrum at 290 K. Although not plotted, negligible overlap also occurs for higher order modes $m>6$. Physically, this small overlap indicates that little of the emitted radiation from a core lens system of size $k_0b=1.8$ will be absorbed by a core lens system of a size parameter $k_0b=2.6$ and vice versa. 

We thus arrive at the principal result of our study. Nanowires surrounded by HMM lenses interact with radiation primarily at a particular wavelength and angular mode with absorptivity that can reach the single channel limit. Therefore, radiation emitted by a nanowire with a certain HMM lens can only minimally exchange radiative heat with other identical nanowires surrounded by lenses of different size parameters. Unlike other selective heating schemes based on plasmonic resonances \cite{Ingvarsson2007,Schmidt2008,Coppens2013,Huang2014,Herzog2014}, the selective resonance identified here is based both on wavelength and angular mode number, enabling high selectivity. This effect is harder to realize with the plasmonic resonances of the bare nanowire alone because achieving similar mode selectivity close to the single channel limit requires tuning both size parameter or material permittivity of the nanowires, while all material properties remain fixed with our core-lens system.

We investigate the origin of the angular selectivity by comparing the observed resonance with previous applications of curvilinear HMMs as hyperlenses \cite{Jacob2006,Liu2007}. Hyperlenses are used to convert high angular momentum, evanescent modes to propagating modes using conservation of angular momentum as the mode propagates radially outward. The mode becomes propagating inside the HMM lens when size parameter $k_0b\geq m$. However, $k_0b=1.8$ for the $m=4$ mode on resonance in Fig.~\ref{fig:mode_analyze}(a), indicating that the excitation in vacuum is actually evanescent. This observation indicates that the HMM lens here is modifying the plasmon resonance of the core similar to the mechanism of enhancement in Ref. 43 rather than converting evanescent and propagating waves. We confirm that the resonance is plasmonic in nature by noticing that little absorption is observed for the polarization for which $\mathbf{E} \parallel z$.  

The origin of the selectivity is also not solely due to the hyperbolic dispersion. HMMs are typically of interest because the hyperbolic dispersion occurs over a broad spectral range, as is the case here. However, Fig. 2(a) and the inset shows that the mode selectivity only occurs around the $\epsilon_{\theta}$ close to zero region of the HMM dispersion, making the selectivity narrowband. The angular selectivity thus requires the anisotropic properties of the HMM but also the epsilon-near-zero (ENZ) region of the dispersion along the $\theta$ direction.

Next, we examine the angular mode selectivity using the well-known single channel limit for absorption and scattering. Physically, the single-channel limit is achieved when radiative damping and absorptive loss both contribute equally to the absorption efficiency of the mode \cite{Hamam2007,Ruan2010,Mann2013}. Mathematically, from Equation \ref{Eq:Qabs} the maximum partial absorption cross-section occurs \cite{Schuller2009} when $\operatorname{Re}(a_m)=1/2$ and $\operatorname{Im}(a_m)=0$, yielding $Q_{abs,m}=1/(2 k_0 a)$. For example, when $a=0.1\lambda$, the limit for partial emissivity is $Q_{abs,m}\approx0.796$ as indicated in Fig.~\ref{fig:mode_analyze}(b). Figure \ref{fig:wavelength_dept}(c) plots the real and imaginary part of the coefficient $a_m$ in Equation \ref{Eq:field} for mode $m=4$ demonstrating that this mode meets the conditions required to reach the single-channel limit for $k_0b=1.8$. Likewise, modes $m=5$ and $m=6$ reach the single-channel limit in Fig.~\ref{fig:mode_analyze}(b) and satisfy the same conditions for $a_m$ at their respective resonant size parameters. However, due to the wavelength-dependence of permittivity, the requirements of the single-channel limit for a fixed size parameter can be met for a single angular mode but are unlikely to be satisfied for other angular modes, as in Fig.~\ref{fig:wavelength_dept}(a). This sensitivity of the angular resonance to the conditions of the single-channel limit contributes to the mode selectivity. 

\begin{figure}[h!]
\centering
\includegraphics[width=1\textwidth]{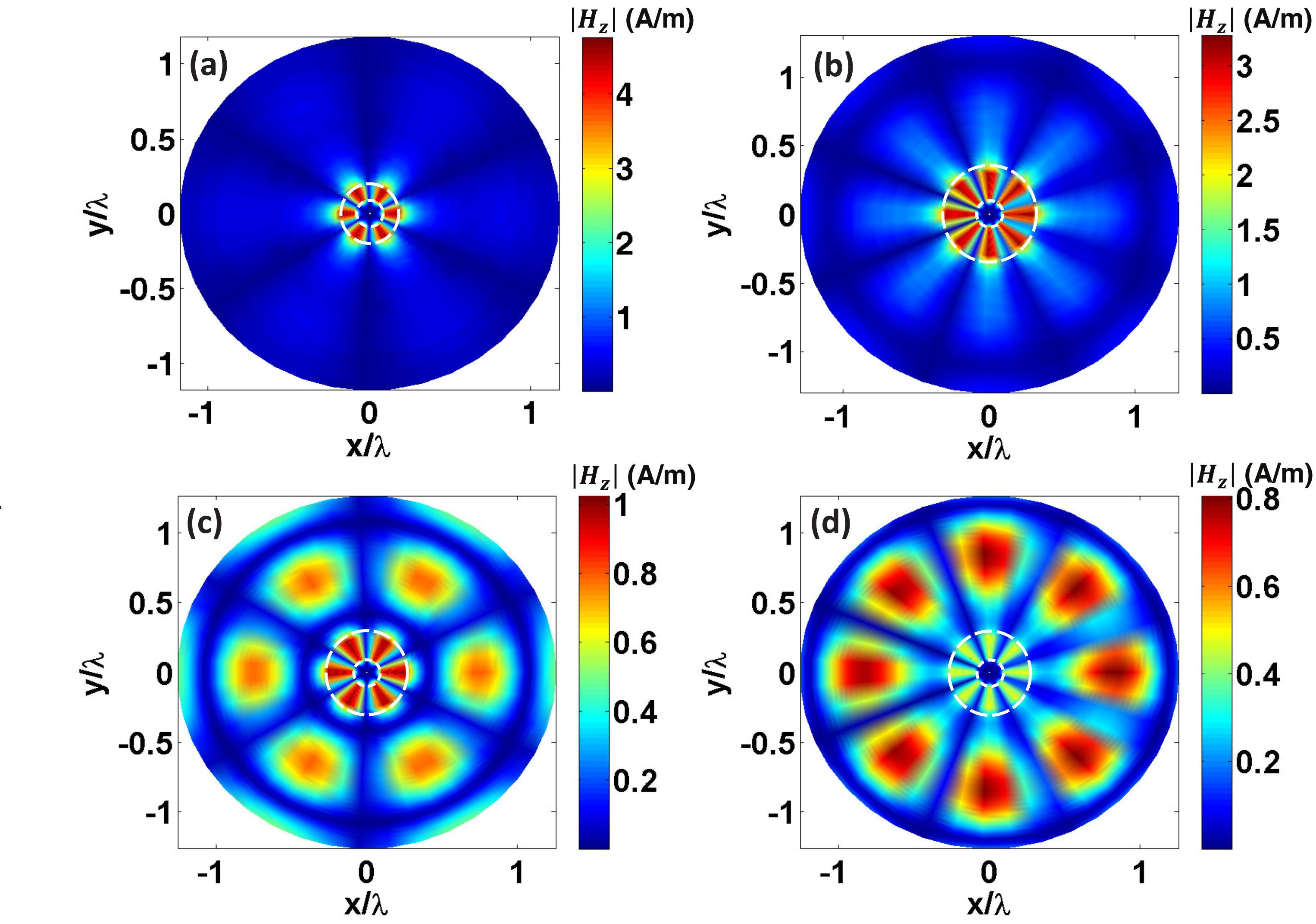}
\caption{\label{fig:field_profile} Field magnitude $\lvert H_z \rvert$ plotted versus x and y coordinates normalized by wavelength, of mode $\lvert m \rvert$  for the EMT-md case in Fig.~\ref{fig:mode_analyze}(a). (a) $\lvert m \rvert=3$, $k_0 b\approx1.1$. (b) $\lvert m \rvert=4$, $k_0 b\approx1.9$. (c) $\lvert m \rvert=3$, $k_0 b\approx1.62$ and (d) $\lvert m \rvert=4$, and $k_0 b\approx1.62$. The dashed white circles represent the approximate inner and outer boundaries of the lens. (a) and (b) are at size parameters of resonances in Fig.~\ref{fig:mode_analyze}(a) and we observe a dominant confined single mode with high field magnitude. However, (c) and (d) correspond to an off-resonant size parameter in which both modes are not confined and have lower field magnitudes than (a) and (b).}
\end{figure}

We further investigate this modal selectivity by examining the resonant mode profiles using the TMM calculation. We reconstruct the field profile of $\lvert H_z \rvert$ in 2D for each mode $\lvert m \rvert$ with incident plane wave direction defined in Fig.~\ref{fig:mode_analyze}(a). Although $a_m$ is symmetric for positive and negative $m$, we must account for the phase factors $\exp{(im\phi)}$ to accurately plot the spatial profile. Figure \ref{fig:field_profile} (a-d) show the 2D plots of $\lvert H_z \rvert$ corresponding to three different size parameters in Fig.~\ref{fig:mode_analyze}(a) for the TMM-md case. The field magnitude $\lvert H_z \rvert$ at resonant size parameters $k_0 b\approx1.1$ ($\lvert m \rvert=3$) and $k_0 b\approx1.9$ ($\lvert m \rvert=4$) are plotted in Fig.~\ref{fig:field_profile}(a) and (b), respectively. In Fig.~\ref{fig:field_profile}(c) and (d), we plot $\lvert H_z \rvert$ for modes $\lvert m \rvert=3$ and $\lvert m \rvert=4$, respectively, for an intermediate size parameter $k_0b\approx1.62$ that is off resonance. We observe from Fig.~\ref{fig:field_profile}(a) and (b) that the lobe patterns at the resonant mode number are highly-confined within the HMM lens. In contrast, in Figs.~\ref{fig:field_profile}(c) and (d) the modes are not confined. Additionally, the fields magnitudes $\lvert H_z \rvert$ in Fig.~\ref{fig:field_profile}(a) and (b) are higher than in Fig.~\ref{fig:field_profile}(c) and (d) by a factor between 3 to 4. The strong, localized field intensities in Fig.~\ref{fig:field_profile}(a) and (b) highlights the modal selectivity of the resonances at specific size parameters. 

We can gain further insight into the origin of the thermal emission spectrum by examining the bulk behavior of an equivalent planar structure. We use the planar Transfer Matrix Method (pTMM) to simulate the equivalent bulk HMM structure on a semi-infinite metallic substrate of the same permittivity of $-1.05+0.01i$ as the core in Fig.~\ref{fig:mode_analyze}. The HMM has the same bi-layer thickness of $\lambda/400$ and material arrangement, including the air-gap, as the TMM-md case of the HMM lens calculation in Fig.~\ref{fig:mode_analyze}. We relate the wave vector component parallel to the vacuum-HMM interface $k_{\parallel}$ in the planar case to $m$ in the cylindrical case by approximating the mode to lie within the HMM \cite{Ruan2010} such that $k_{\parallel}=m/r_{eff}$ where $r_{eff}=(a+b)/2$. The penetration of the modes through the HMM to the absorbing layer can be observed by the non-zero imaginary part of the Fresnel reflection coefficient $\operatorname{Im}(R_p)$ which describes the absorption of the incident evanescent field \cite{Kidwai2012}.  

\begin{figure}[h!]
\centering

    \includegraphics[width=1\textwidth]{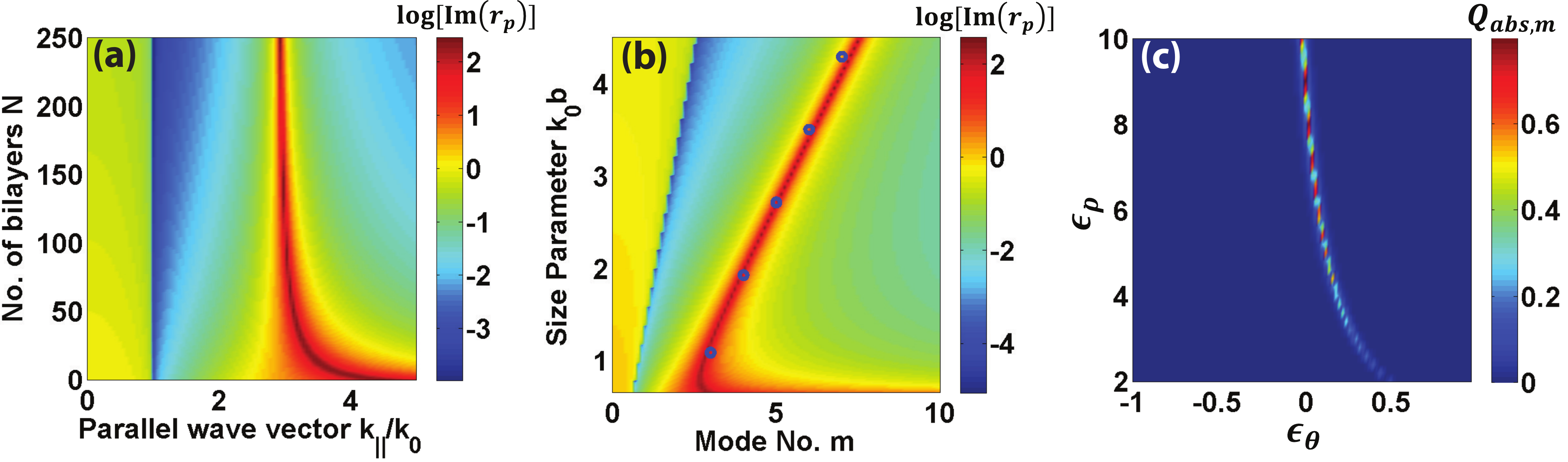}
\caption{\label{fig:param_sweep} (a) Log plot of the imaginary part of the Fresnel reflection coefficient $\operatorname{Im}(R_p)$, indicating the magnitude of absorption of the incident evanescent field, using pTMM for different values of $k_{\parallel}/k_0$ and number of metal-dielectric bi-layers $N$. The HMM lowers the parallel momentum required for the resonance with slow variation versus number of bi-layers. (b) $\log[\operatorname{Im}(R_p)]$ for the planar case in (a) compared to the peak positions of the TMM-md case (symbols) in Fig.~\ref{fig:mode_analyze}(b) for different equivalent values of $m$ and size parameter $k_0b$. The agreement between the planar and cylindrical calculations indicates that the composite plasmonic resonances are of the same nature. (c) Partial emissivity $Q_{abs,m}$ for $m=4$ mode at a size parameter of $k_0b\approx1.8$ for EMT-HMM case in Fig.~\ref{fig:mode_analyze}(a) for different values of $\epsilon_{\rho}$ and $\epsilon_{\theta}$. The region of interest for selective heating is $\epsilon_{\rho} \gtrsim 5,\epsilon_{\theta}<0$ for which the emissivity of the resonant mode is largest. }
\end{figure}

We plot $\log[\operatorname{Im}(R_p)]$ obtained from pTMM against the normalized parallel wave vector $k_{\parallel}/k_0$ and number of HMM bi-layers in Fig.~\ref{fig:param_sweep}(a). As $N$ increases, the position of maximum $\operatorname{Im}(R_p)$ decreases from the metal-vacuum surface plasmon condition of $k_{\parallel}/k_0\approx4.6$ to around $k_{\parallel}/k_0\approx3$, decreasing the high parallel momentum for plasmonic resonance when the HMM is present. As $m$ is a measure of the angular momentum, the above relationship $k_{\parallel}=m/r_{eff}$ indicates that the angular momentum for the mode is reduced, for a fixed effective radius $r_{eff}$, when $k_{\parallel}$ is decreased. We also plot $\log[\operatorname{Im}(R_p)]$ versus the converted effective $m$ and size parameter $k_0b$ in Fig.~\ref{fig:param_sweep}(b) and overlay the positions of the resonances of the cylindrical case in Fig.~\ref{fig:mode_analyze} onto Fig.~\ref{fig:param_sweep}(b). The resonant peaks in the cylindrical case closely follow the prediction of the planar case, allowing us to conclude that both resonances are of the same nature. 

From this planar analysis, we can understand the relationship between the size parameter and mode number of the resonances in Fig.~\ref{fig:mode_analyze}(b). After approximately 50 bi-layers, the parallel momentum required to excite the resonance becomes nearly constant as in Fig.~\ref{fig:param_sweep}(a). From the relation $k_{\parallel}=2m/(a+b)$, if $k_{\parallel}$ is constant as $b$ increases $m$ must also increase, leading to the nearly linear increase of the mode number with size parameter as in Fig.~\ref{fig:mode_analyze}(b). 
 
We now examine the optical properties of the HMM lens and core that will allow the selectivity by studying how the partial emissivity of a mode depends on the permittivity of the HMM lens. Figure \ref{fig:param_sweep}(c) plots the partial emissivity for the $m=4$ mode ($k_0b\approx1.8$ for EMT-HMM case in Fig.~\ref{fig:mode_analyze}(a)) as $\epsilon_{\rho}$ and $\epsilon_{\theta}$ varies. From Fig.~\ref{fig:param_sweep}(c), the region of $\epsilon_{\rho} \gtrsim 5$ and a negative but close to zero value of $\epsilon_{\theta}$ is where the largest enhancement occurs. The enhancement for these permittivity values can be explained by the dispersion relation in the HMM \cite{Jacob2006}, $k_{\rho}^2/|\epsilon_{\theta}|=k_{\theta}^2/\epsilon_{\rho}-k_0^2$, and noting that small and negative $\epsilon_{\theta}$, with $k_{\theta}/k_0\approx 3$ and $\epsilon_{\rho}=10$ for example, causes $k_{\rho}$ to be very small and imaginary and allows the field to extend to the inner absorbing core. The sensitivity of the mode selective plasmonic resonances to the HMM parameters is unlike typical broadband enhancement effects of HMMs \cite{Biehs2012,Guo2013}.

\begin{figure}[h!]
\centering
\includegraphics[width=.45\textwidth]{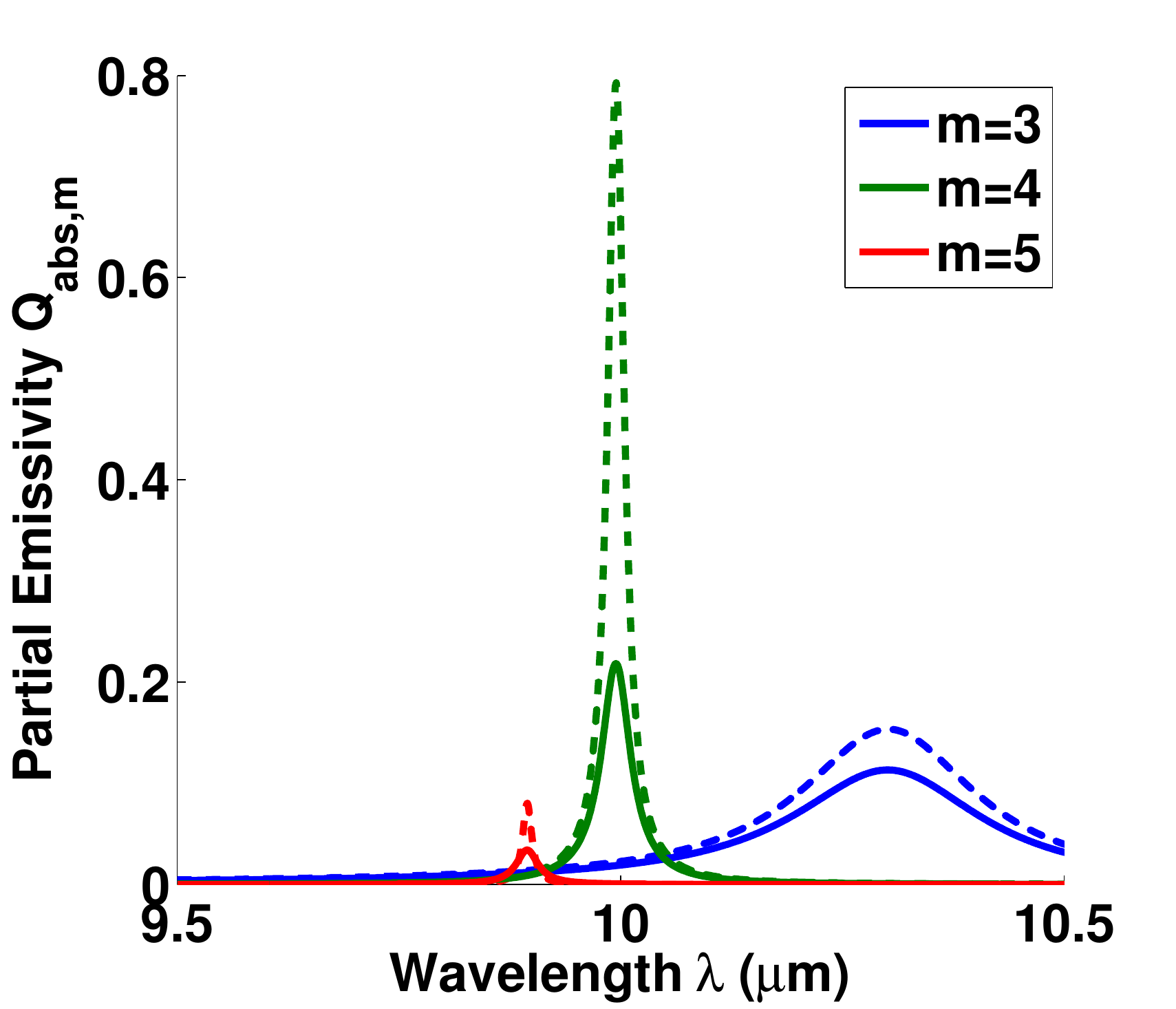}
\caption{\label{fig:wavelength_dept_loss}  Partial emissivity $Q_{abs,m}$ versus wavelength for $m=3,4,5$ with loss (solid lines) and without loss (dashed lines) in the HMM lens (where $a=1\ \mu$m and $k_0 b=1.8$ for the mode $m=4$ in Fig.~\ref{fig:mode_analyze}). The presence of loss in the lens decreases the resonant absorption peak, $m=4$, while the difference in emissivity between off-resonant modes such as $m=3$ and the resonant $m=4$ mode decrease. The colors indicating mode number $m$ are the same for $Q_{abs,m}$ with and without loss.}
\end{figure}

Finally, we consider the effect of loss in the HMM lens. Physically, loss results in the lens also playing a role in the radiative transfer. Since the temperature of the lens is not fixed, HMM lens will equilibrate to a temperature close to that of the heated core, allowing us to consider the core-lens structure as a single object for the purposes of analyzing radiative emission. We incorporate loss by modifying the Drude dispersion of the metal to have $\gamma=0.001\omega_p$ so that $(\epsilon_m,\epsilon_d)=(-5.1 + .015i,3.4)$ at $10\ \mu$m. The partial emissivity $Q_{abs,m}$ is now defined with respect to the size of the whole structure $b$. As shown in Fig.~\ref{fig:wavelength_dept_loss}, adding loss decreases the peak absorptivity around $10\ \mu$m for $m=4$ compared to the lossless case of Fig.~\ref{fig:wavelength_dept}(a). Also, loss decreases the relative differences between absorption of an adjacent peak that is of a different angular mode. We conclude that the loss reduces the angular mode and wavelength selectivity for selective heating and thus that fully exploiting the thermal HMM lens requires low-loss plasmonic materials in the infrared. Materials such as 6H-SiC have shown potential for low-loss with negative real permittivity in the mid-infra-red range \cite{Caldwell2013}.
 
In summary, we have theoretically demonstrated a new approach to selective radiative heating based on tuning angular mode resonances with HMM lenses. This approach enables high selectivity for radiative exchange due to the requirement that both wavelength and angular mode number of the emitter and absorber match for localized radiative heat transfer. Our result could have applications in radiative thermal management.

\begin{acknowledgments}
This work is part of the 'Light-Material Interactions in Energy Conversion' Energy Frontier Research Center funded by the U.S. Department of Energy, Office of Science, Office of Basic Energy Sciences under Award Number DE-SC0001293. D.D. gratefully acknowledges fellowship support from the Agency for Science, Technology and Research (Singapore). 
\end{acknowledgments}

\nocite{*}
%\bibliographystyle{C:/texlive/2013/texmf-dist/bibtex/bst/revtex4/apsrev-title.bst} 
%\bibliography{ref_v21.bib}

\end{document}